# Slowdown and Heating of Interstellar Neutral Helium by Elastic Collisions Beyond the Heliopause

P. Swaczyna[1,*], F. Rahmanifard[2], E. J. Zirnstein[1], D. J. McComas[1], J. Heerikhuisen[3]

[1]Department of Astrophysical Sciences, Princeton University, Princeton, NJ 08544, USA
[2]Physics Department, Space Science Center, University of New Hampshire, Durham, NH 03824, USA
[3]Department of Mathematics and Statistics, University of Waikato, Hamilton, New Zealand
[*]Corresponding author (swaczyna@pricenton.edu)

**Abstract**

Direct sampling of interstellar neutral (ISN) atoms close to the Sun enables studies of the very local interstellar medium (VLISM) around the heliosphere. The primary population of ISN helium atoms has, until now, been assumed to reflect the pristine VLISM conditions at the heliopause. Consequently, the atoms observed at 1 au by the Interstellar Boundary Explorer (IBEX) were used to determine the VLISM temperature and velocity relative to the Sun, without accounting for elastic collisions with other species outside the heliopause. Here, we evaluate the effect of these collisions on the primary ISN helium population. We follow trajectories of helium atoms and track their collisions with slowed plasma and interstellar hydrogen atoms ahead of the heliopause. Atoms typically collide a few times in the outer heliosheath, and only ~1.5% of the atoms are not scattered at all. We use calculated differential cross sections to randomly choose scattering angles in these collisions. We estimate that the resulting primary ISN helium atoms at the heliopause are slowed down by ~0.45 km s$^{-1}$ and heated by ~1100 K compared to the pristine VLISM. The resulting velocity distribution is asymmetric and shows an extended tail in the antisunward direction. Accounting for this change in the parameters derived from IBEX observations gives the Sun's relative speed of 25.85 km s$^{-1}$ and temperature of 6400 K in the pristine VLISM. Finally, this paper serves as a source of the differential cross sections for elastic collisions with helium atoms.

## 1. Introduction

Neutral components of the interstellar matter penetrate heliospheric boundaries and are used to diagnose the very local interstellar medium (VLISM) in front of the heliosphere. Comprehensive observations of interstellar neutral (ISN) helium with the IBEX-Lo sensor (Fuselier et al. 2009) on the Interstellar Boundary Explorer (IBEX, McComas et al. 2009) mission reveal the VLISM temperature and its relative velocity to the Sun (Möbius et al. 2009b, 2009a). Analyses of these observations have assumed that the main population of ISN helium is not modified outside the heliopause, and therefore the atom fluxes measured at 1 au can be directly related to their distribution function in the pristine VLISM, i.e., before it is affected by the heliosphere, using Liouville's theorem (Lee et al. 2012, 2015; Sokół et al. 2015). Additionally, atoms are ionized in the inner heliosphere due to charge exchange with solar wind ions, electron impact ionization, and photoionization (Sokół et al. 2019). These analyses showed that the ISN helium temperature is 7500 K and flows with a speed of 25.4 km s$^{-1}$ relative to the Sun toward direction (75.7°, −5.1°) in the ecliptic coordinates (Bzowski et al. 2015; McComas et al. 2015; Schwadron et al. 2015). These parameters are often interpreted as the pristine relative velocity and temperature of the VLISM because the primary ISN helium so far has not been expected to be significantly modified beyond the heliopause in the region perturbed by the heliosphere (the outer heliosheath).

The Warm Breeze discovery indicated that another ISN helium population also exists in the outer heliosheath (Kubiak et al. 2014, 2016) with a higher temperature and lower speed. Bzowski et al. (2017, 2019) showed that this population is created from He$^+$ ions in the outer heliosheath that charge exchanged with the primary population of the ISN helium. Both populations are detected in the three lowest energy steps of IBEX-Lo (Swaczyna et al. 2018). Still, even with a comprehensive uncertainties system (Swaczyna et al. 2015), the observations show statistically significant differences from the predictions. This discrepancy may suggest that the model is too simple to fully represent the observations. Swaczyna et al. (2019a) suggested that the primary ISN helium is not fully equilibrated in the VLISM and, instead of a Maxwellian, follows a kappa distribution. Additionally, Wood et al. (2019) showed that a bi-Maxwellian distribution with different perpendicular and parallel temperatures ($T_\perp/T_\parallel = 0.62$) relative to the inflow direction gives a significant improvement in the goodness of fit.

In the outer heliosheath, charge exchange collisions between helium atoms and He$^+$ ions dominate all other ionization processes of helium atoms. The resonant character of this process is often used to claim that the momentum of the newly created atom is the same as the momentum of the parent ion. Swaczyna et al. (2019b) demonstrated that this simplification is not justified for typical collision speeds in the outer heliosheath, and angular scattering of colliding particles impacts the distribution function of the created neutral population. However, this angular scattering can be neglected in studies of the global heliosphere because the amount of momentum exchange is not significantly changed (Heerikhuisen et al. 2009; Izmodenov et al. 2000).

In this study, we focus on elastic collisions of helium atoms that are rarely considered in the heliospheric studies. While elastic collisions do not change charge state of particles, they lead to momentum exchange. The heating of interstellar atoms due to such collisions with the solar wind has been considered since the 1970s. Many of these studies (Fahr 1978; Fahr et al. 1985; Holzer 1977; Kunc et al. 1983; Wallis 1975) applied a continuous momentum transfer approach and concluded that this process might increase the temperature of ISN atoms at 1 au by a few hundred to a few thousand kelvins from the interstellar value. However, Gruntman (1986) argued that this approach should not be used as each atom undergoes a small number of binary collisions inside the heliosphere, and thus the continuous transfer approximation is not valid. Additionally, Chassefière et al. (1986) pointed out that Fahr et al. (1985) significantly overestimated the elastic cross section, consequently overemphasizing heating by elastic collisions with the solar wind. Recently, Gruntman (2013, 2018) used a "one or none" collision approximation for this process and showed that they increase ISN helium temperature by ~200 K and slightly enhance the tail of the distribution due to rare but large scattering angles.

Chassefière and Bertaux (1987) considered elastic collisions outside the heliopause and concluded that about 10–30% of ISN helium atoms are elastically scattered to form an additional slowed (by ~3 km s$^{-1}$) and heated (by ~4000 K) population. This population should not be confused with the secondary population created from the charge exchange collisions. They obtained this result using a momentum transfer cross section that significantly underestimates the number of collisions in the outer heliosheath but simultaneously increases momentum exchange in these collisions. In the present study, we revisit this problem and present how these collisions change the interpretation of IBEX observations.

## 2. Methodology

Far from the heliopause, charged and neutral populations in the pristine VLISM are in thermal equilibrium and thus can be described by Maxwell distributions. However, in the outer heliosheath, the interstellar plasma diverges to flow around the heliopause, while ISN atoms can still penetrate the heliosphere (Axford 1972; Parker 1961; Zank 1999). Consequently, flows of ions and atoms are separated, so they are no longer in thermal equilibrium. While elastic collisions are not frequent enough to restore equilibrium, they affect the neutral components passing the outer heliosheath. In this study, we estimate this effect using Monte Carlo calculations.

*2.1 Elastic Collision Scattering*

Calculations of elastic collision scatterings require using relevant differential cross sections. The order of magnitude of integral cross sections for various elastic scatterings is similar. Hence, we consider the four most abundant species in the VLISM: hydrogen atoms, protons, helium atoms, and $He^+$ ions. Collisions with less abundant species can be safely neglected due to orders of magnitude lower abundances in the VLISM (Frisch et al. 2011).

Separation of charge exchange and elastic scatterings for collisions of ions and atoms of the same species (e.g., $He^+ - He^0$) is not possible from quantum mechanical considerations because the colliding nuclei are not distinguishable. However, differential cross sections calculated in the indistinguishable approach show two distinctive maxima at scattering angles between directions of the incoming and outgoing atom close to 0° and 180° (Barata & Conde 2010; Krstić & Schultz 1998; Schultz et al. 2016). The scattering amplitudes contributing to each maximum can be separated and used to calculate two separate elastic and charge exchange cross sections. This approximation does not account for interference of scattering amplitudes, but this effect can be neglected for collision speeds considered in this study. In this study, we follow the distinguishable particle approach (Appendix A). Moreover, we do not use the momentum transfer cross sections that can be used as an effective term for a continuous momentum transfer between populations because such cross section cannot be used to track individual collisions between particles.

Comprehensive sources of these cross sections are limited. Krstić & Schultz (1998) calculated differential cross sections for collisions of helium atoms with protons, which are available in the ALADDIN database[*] in a tabulated form for 24 center-of-mass energies in the range 0.1 – 100 eV. We do not find any other comprehensive set of the needed differential cross sections. Therefore, we calculate these cross sections using the JWKB method as described by Nitz et al. (1987), utilizing interaction potentials for collisions of helium atoms with hydrogen atoms (Gao et al. 1989), protons (Helbig et al. 1970), helium atoms (Ceperley & Partridge 1986), and $He^+$ ions (Barata & Conde 2010; Marchi & Smith 1965). The calculations are performed for a wide range of energies from $10^{-4}$ to $10^4$ eV, as described in Appendix A. We check that our calculations for collisions with protons are consistent with the differential cross sections obtained by Krstić & Schultz (1998). Figure 1 shows the obtained elastic collision cross sections of helium atoms for energies of 0.01, 0.1, 1, and 10 eV to span over the collision energy range expected in the outer heliosheath. Atom–ion differential cross sections show characteristic oscillations related to more partial wave contributions, as phase shifts decrease much more slowly with the order of the partial wave than in the case of atom–atom collisions. We check the obtained cross sections against previously obtained sources for the available energies and collision pairs (Barata & Conde 2010; Gao et al. 1989; Krstić & Schultz 1998). As the collision energy increases, the scattering angles in elastic collisions are smaller.

---

[*] https://www-amdis.iaea.org/ALADDIN/

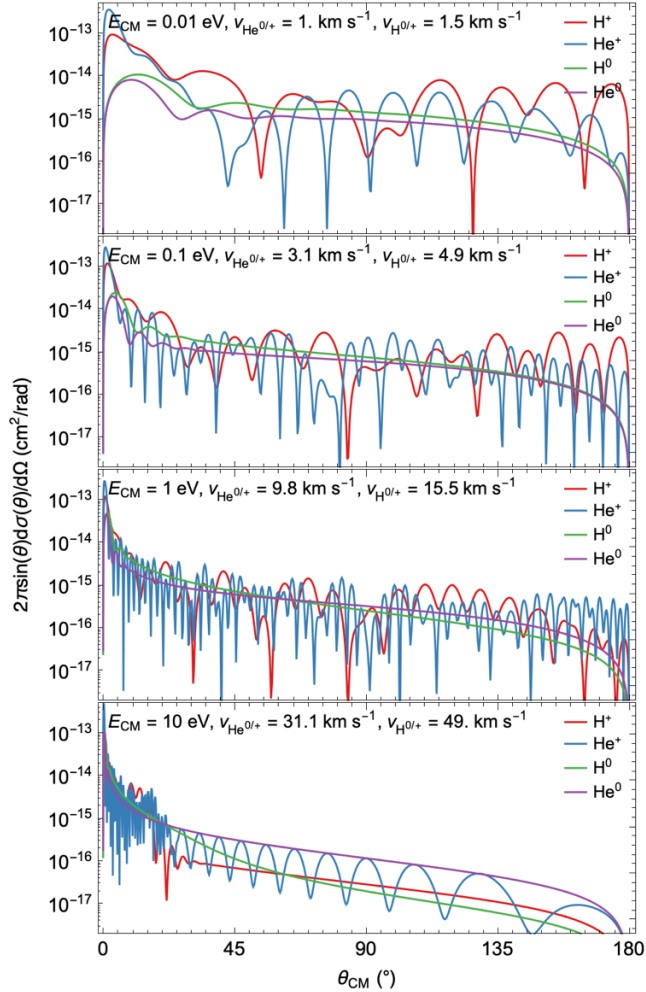

**Figure 1** – Differential cross section for elastic collisions of helium atoms with protons (red line), He$^+$ ions (blue), hydrogen atoms (green), and helium atoms (purple). Three panels represent collision energies 0.01, 0.1, 1, and 10 eV (top to bottom).

We estimate the scattering effect on the relative motion of helium atoms colliding with these species, as presented in Figure 2. Panel (a) shows the integral cross section for these collisions as a function of relative speed. Collisions with protons and He$^+$ ions show stronger dependence on the relative speed, but all are comparable for speeds relevant for this problem. Mean scattering angles, presented in panel (b), are significantly larger at lower speeds. Panel (c) displays the mean relative change of the speed in the helium atom initial motion direction. This change is between ~0.05 and 1.5 km s$^{-1}$ per collision and depends on colliding species and relative speed. Such changes are not negligible compared to the bulk flow of the VLISM relative to the Sun. Moreover, all considered elastic cross sections are higher than the charge exchange cross section with He$^+$ ions, thus we expect that these collisions affect a significant portion of the helium atom population. For this paper, we only consider the consequences of more frequent elastic collisions. However, charge exchange collisions, mostly with He$^+$ ions, result in creating the Warm Breeze and remove some atoms from the primary population (Bzowski et al. 2019), but this effect is not considered in this study. Consequences for the primary ISN helium should be much smaller than found for the primary ISN hydrogen by Pogorelov et al. (2008, 2009) due to lower ionization rate and narrower distribution of helium atoms in the velocity space.

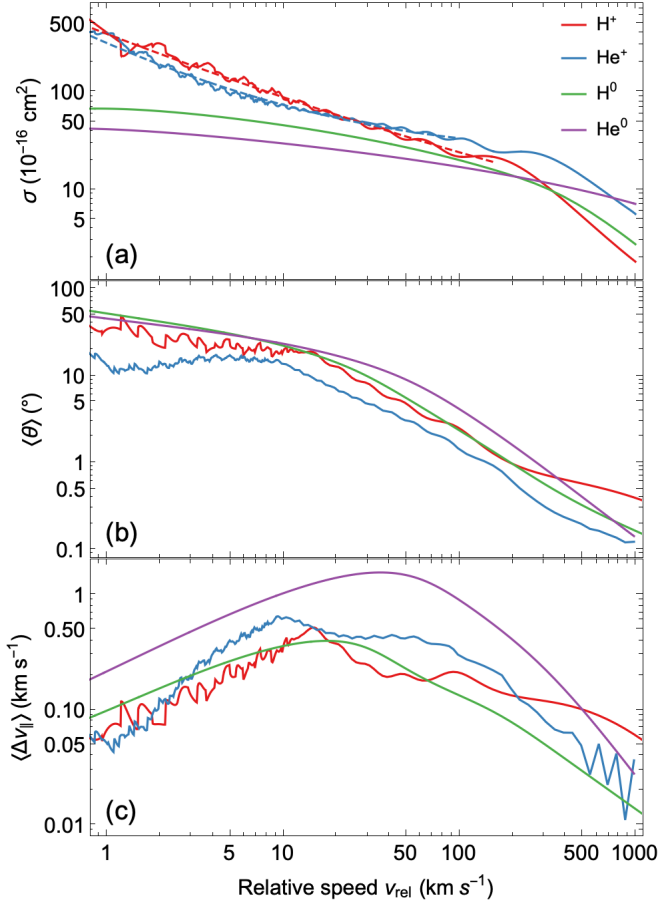

**Figure 2** – Elastic collision characteristics as a function of the relative speed of helium atoms with protons (red line), He$^+$ ions (blue), hydrogen atoms (green), and helium atoms (purple): (a) integral cross section, (b) mean scattering angle, (c) mean parallel speed change per collision. The dashed line in panel (a) shows a parabolic fit in log–log scale for collision energies below 100 eV (see Appendix B)

*2.2 Outer Heliosheath Plasma and Neutrals*

ISN atoms, observed by IBEX at 1 au, enter the heliosphere in the proximity of the heliospheric nose (Kubiak et al. 2014). To estimate the first-order effects of elastic scattering of the primary population of ISN helium atoms, we employ a one-dimensional cut of a global numerical model of the heliosphere (Heerikhuisen et al. 2015; Zirnstein et al. 2016) along the inflow direction. This model couples the MHD description of plasmas with kinetic transport and charge-exchange collisions of neutrals. The model provides bulk densities, speeds, and temperatures of the plasma and two populations of ISN hydrogen atoms. The primary and secondary populations consist of atoms originating from the pristine VLISM, and the outer heliosheath, respectively. The plasma is described using a single population of protons. However, the interstellar plasma also contains a non-negligible population of helium ions, and we want to account for collisions of helium atoms with protons and helium ions separately. Following Bzowski et al. (2019), we use the same bulk flow velocities and temperatures for protons and He$^+$ ions taken from the model and assuming their densities are in a constant ratio everywhere in the outer heliosheath. They found the proton density of 0.054 cm$^{-3}$ and the He$^+$ density of 0.009 cm$^{-3}$ at the outer boundary of the model, which yields a total plasma density of 0.09 nuc cm$^{-3}$.

## 2.3 Transport of Neutral Helium Atoms

In the one-dimensional cut described in Section 2.2, the heliopause is located at 110 au from the Sun, and a significant deviation from the pristine VLISM conditions extends up to ~350 au. We calculate evolution of the velocity distribution function of helium atoms due to elastic collisions from the pristine VLISM to the heliopause. While the parameters of plasma and neutrals in this cut depend only on the one-dimensional distance from the heliopause, we track helium atom trajectories in three dimensions. The inflow direction is aligned with the *z*-axis of the Cartesian coordinate system used in this study. For consistency, we neglect a small flow perpendicular to the *z*-axis, and thus the problem has rotational symmetry in the *xy*-plane. Details of the Monte Carlo calculations are presented in Appendix B.

Based on these calculations, we investigate samples of the distribution functions in 10-au wide bins along the inflow direction from the pristine VLISM to the heliopause. As ISN helium atoms move through the outer heliosheath, their distribution function starts to deviate from the Maxwell distribution significantly. We find that they can be well described by an asymmetric version of the kappa distribution. In the direction perpendicular to the inflow direction, it follows the standard formulation of the kappa distribution (Livadiotis & McComas 2013), while in the parallel direction, it is best characterized by a composition of two kappa distributions, with different parameters below and above the peak velocity. A mathematical formulation of this distribution is presented in Appendix C.

Additionally, we fit the Maxwell distribution function to the part of the distribution close to the peak, defined by where the probability distribution function is at least 5% of the maximum. This selection is a typical range of ISN fluxes interpreted from the IBEX-Lo observations (e.g., Bzowski et al. 2015). We note that the fit parameters do not necessarily match the distribution moments of the helium atom velocities due to a significant departure from the Maxwell distribution.

## 3. Results

Following the method presented in Section 2.3, we investigate changes in the ISN helium distribution function in the outer heliosheath. The pristine VLISM populations follow the Maxwell distribution with the speed $u_{\text{VLISM}} = 25.4$ km s$^{-1}$ relative to the Sun, and the temperature $T_{\text{VLISM}} = 7500$ K. Figure 3 shows the resulting distribution function of the primary atoms at the heliopause. The probability distribution function is widened because of elastic collisions in the outer heliosheath but shows significant asymmetry in parallel and perpendicular directions to the inflow direction (*z*-axis). Moreover, the parallel component shows an extended tail in the anti-sunward direction. The average parallel component of the atoms' velocities is $u_{\text{HP}} = 24.85$ km s$^{-1}$, and their temperature is $T_{\text{HP}} = 8850$ K. The temperature describes the mean energy of the particles and is related to the mean square speed $\langle v^2 \rangle$ as $T = \frac{m \langle v^2 \rangle}{3 k_{\text{B}}}$, where $m$ is the atomic mass, and $k_{\text{B}}$ is the Boltzmann constant.

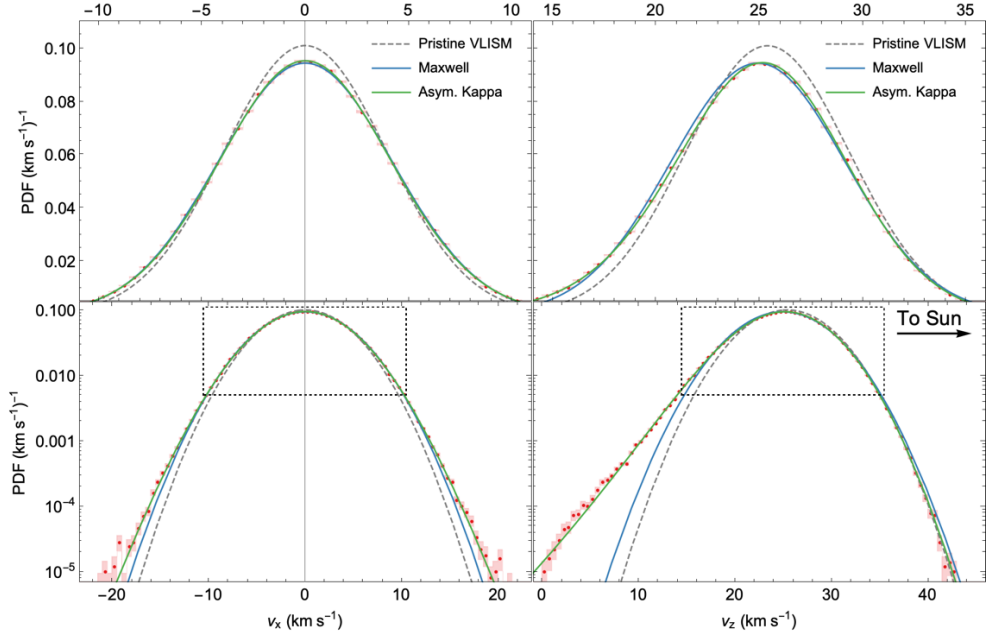

**Figure 3** – Probability distribution function of the primary ISN helium at the heliopause. The left and right panels show distributions in perpendicular and parallel directions to the inflow direction, respectively. The top panels show part of the distribution marked with dotted boxes in linear scale. The calculated distribution is marked as red histograms with boxes showing bin sizes and method uncertainties. Blue and green lines show the fitted Maxwell and asymmetric kappa distributions. The initial distribution is marked with dashed gray lines.

The fitted Maxwell distribution has the relative speed $u_M = 24.95$ km s$^{-1}$, and the temperature is $T_M = 8600$ K. This function well describes the perpendicular component of the distribution, with only small discrepancies in the tails, but fails to describe the asymmetric parallel component. The peak of the Maxwell fit is shifted to lower relative speed than the peak, and the extended anti-sunward tail is not captured. However, the asymmetric kappa distribution introduced in Section 2.3 reproduces all these features. The peak position is at $u_\kappa = 25.15$ km s$^{-1}$. The perpendicular component has temperature $T_{\kappa\perp} = 8650$ K, and kappa index $\kappa_\perp = 30$. This finding shows that this component is very close to the Maxwell distribution, and only small differences are expected in the tails of the distribution (see Figure 3). The parallel part has different parameters for the sunward ($T_{\kappa\|S} = 7950$ K, $\kappa_{\|S} = \infty$), and anti-sunward ($T_{\kappa\|A} = 10600$ K, $\kappa_{\|A} = 6.5$) parts of the distribution. Note that the sunward part is consistent with the Maxwell distribution, which is represented by $\kappa_{\|S} = \infty$.

The evolution of the distribution parameters in the outer heliosheath is presented in Figure 4. The distribution of the primary ISN helium starts to change properties around 240 au from the Sun, i.e., where significant deviations of the plasma and neutrals flow from the pristine VLISM parameters begin in this model. The properties progressively change across the outer heliosheath, but the change is more rapid closer to the heliopause, where the discrepancy between parameters of the ISN helium and the plasma components is the highest.

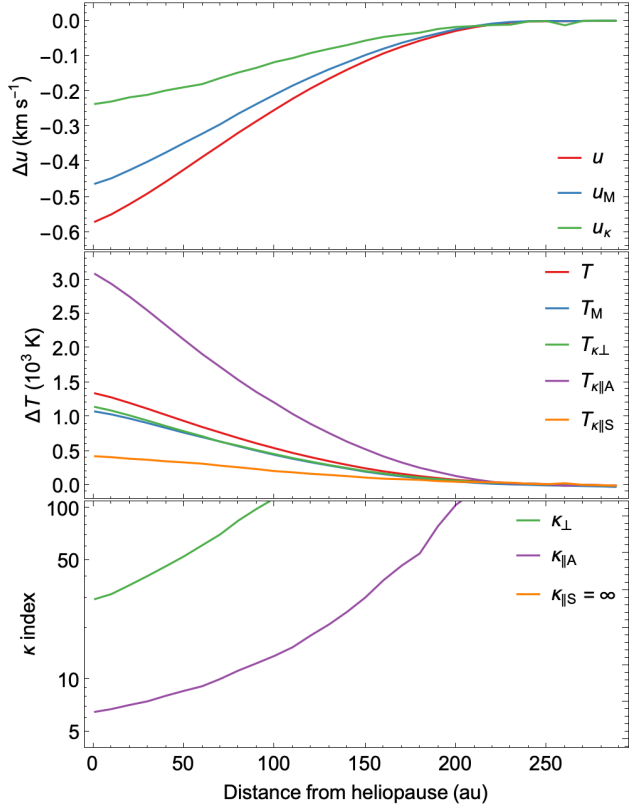

**Figure 4** – Parameters of the calculated distribution function (red) and the fit distributions. The top panel shows the change of the speed in the direction parallel to the inflow direction relative to the pristine VLISM values. The middle and bottom panels present the temperature increases and kappa indices for these distributions. The mean speed and temperature change are shown with red lines. Other colors represent parameters of the fitted Maxwell and asymmetric kappa distribution functions (see the text for details).

The mean numbers of collisions for one helium atom traveling through the outer heliosheath based on the Monte Carlo calculations and their standard deviations are 1.58 ± 1.30, 0.20 ± 0.45, 0.88 ± 1.05, and 1.70 ± 1.32 for collisions with protons, $He^+$ ions, primary ISN hydrogen atoms, and secondary ISN hydrogen atoms, respectively. The average number of all elastic collisions is 4.35, and only 1.5% of primary ISN helium atoms are not elastically scattered in the outer heliosheath at all.

So far, analyses of IBEX observations assumed that the primary ISN helium follows the Maxwell distribution in the outer heliosheath (e.g., Bzowski et al. 2015; Schwadron et al. 2015; Swaczyna et al. 2018). These studies neglected elastic scatterings; therefore, the speed and temperature are different in the pristine VLISM compared to the parameters found previously. To estimate first-order corrections to the parameters of the VLISM derived from the IBEX observations, we assume that the absolute differences in the parameters are the same as in our simulations. Our analysis shows that the speed and temperature at the heliopause following the Maxwell fit change by $\Delta u = u_M - u_{VLISM} = -0.45$ km s$^{-1}$ and $\Delta T = T_M - T_{VLISM} = 1100$ K compared to the pristine VLISM. Assuming that previously derived values correspond to conditions at the heliopause with the Maxwell distribution function $u_{HP,IBEX} = 25.4$ km s$^{-1}$ and $T_{HP,IBEX} = 7500$ K, the pristine VLISM parameters are $u_{VLISM,IBEX} = u_{HP,IBEX} - \Delta u = 25.85$ km s$^{-1}$ and $T_{VLISM,IBEX} = T_{HP,IBEX} - \Delta T = 6400$ K. Furthermore, we expect the ISN helium distribution at the heliopause is better represented by the asymmetric kappa distribution with parameters: $u_{\text{HP},\kappa} = 25.6$ km s$^{-1}$, $T_{\text{HP},\kappa\perp} = 7550$ K, $\kappa_{\text{HP},\perp} = 30$, $T_{\text{HP},\kappa\|\text{S}} = 6800$ K, $\kappa_{\text{HP},\|\text{S}} = \infty$, $T_{\text{HP},\kappa\|\text{A}} = 9500$ K, $\kappa_{\text{HP},\|\text{A}} = 6.5$. To verify if these changes to initial conditions can change our findings, we repeat our calculations with an ad hoc model in which flow speeds are proportionally scaled to match $u_{\text{VLISM,IBEX}}$ at the outer boundary, and all temperatures beyond the heliopause are decreased by $\Delta T$. The obtained relative changes $\Delta u$ and $\Delta T$ agree with these found above.

The changes in the outer heliosheath significantly exceed their uncertainties from previous IBEX studies (Bzowski et al. 2015; Swaczyna et al. 2018). To illustrate how they affect the IBEX-Lo observations, we calculate the expected count rates for three orbits observed in 2009: 14, 16, and 18 for the three distributions characterized above. These rates are calculated using the analytic full integration model (Schwadron et al. 2015). Figure 5 shows results compared with the observed rates on these orbits. The Maxwell distribution with the pristine VLISM parameters shows a significantly narrower signal in spin angle, representing a lower ratio of thermal speed to bulk speed in this case. The asymmetric kappa distribution and the Maxwell fit at the heliopause give similar rates. However, small changes are present in the distribution wings.

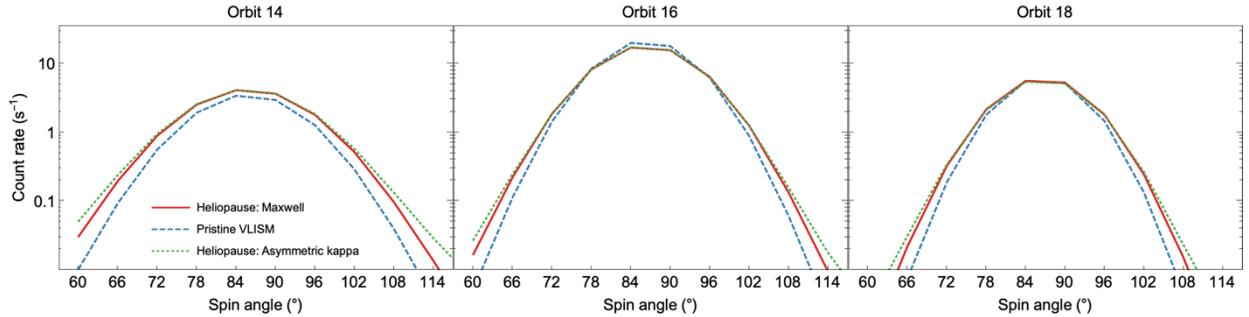

**Figure 5** – Comparison of the expected IBEX-Lo primary ISN helium fluxes for three orbits: 14, 16, and 18 calculated for three distribution functions outside the heliopause (see the text).

## 4. Discussion and Conclusions

The vast majority of primary ISN helium atoms from the VLISM are elastically scattered at least once in the outer heliosheath before entering the heliosphere. Consequently, ISN helium atoms are significantly heated (~1100 K) and slowed (~0.45 km s$^{-1}$), thus their distribution at the heliopause does not represent the pristine VLISM parameters, as is commonly assumed (Möbius et al. 2004). This result shows that the pristine VLISM parameters obtained from IBEX observations need to be adjusted for this effect. Using the current "working values" provided by McComas et al. (2015), we estimate the pristine VLISM relative speed is $u_{\text{VLISM}} \approx 25.4 + 0.45 = 25.85$ km s$^{-1}$, and the temperature is $T_{\text{VLISM}} \approx 7500 - 1100 = 6400$ K. Note that this change is across the IBEX correlation tube of these two parameters (McComas et al. 2012, 2015). Additionally, the estimated heating is about ~5 times stronger than heating due to collisions with the solar wind protons and α-particles, as estimated by Gruntman (2018).

Our results differ significantly from the findings by Chassefière & Bertaux (1987), who estimated that only 10–30% of atoms are elastically scattered and produce an additional component of ISN helium slower by ~3 km s$^{-1}$, and hotter by ~4000 K compared to the primary population. However, the average changes of both populations in their study would yield speed $20\% \times 3 = 0.6$ km s$^{-1}$ and temperature $20\% \times 4000 = 800$ K, which are close to our results. We note that those authors used momentum transfer cross sections that should give correct average changes in momentum but underestimate the number of elastic collisions (Schultz et al. 2016).

The elastic collisions cause a significant asymmetry in the ISN helium distribution at the heliopause naturally aligned with the interstellar bulk flow (see Figures 3 and 4). This alignment is a result of collisions occurring at different positions in the outer heliosheath with slower and heated populations. The magnitude of the temperature change is smaller than the asymmetry found by Wood et al. (2019), who used a bi-Maxwell distribution to model the ISN helium distribution. However, this difference may be caused by two vastly different distribution functions used in the respective studies.

This investigation gives an estimation of the elastic collision effects on the distribution of the ISN helium atoms. Nevertheless, a full quantitative analysis using IBEX observations to determine the pristine VLISM parameters is necessary and requires tracking the elastic scattering across the outer heliosheath using a 3D global model of the heliosphere. Such a study would allow for the determination of possible deflection of the inflow speed in the outer heliosheath that cannot be estimated using the method presented in this paper. Finally, this paper also serves as a source of the cross sections for helium atom elastic collisions in the outer heliosheath.

*Acknowledgments*: This material is based upon work supported by the National Aeronautics and Space Administration under Grant No. 80NSSC20K0781 issued through the Outer Heliosphere Guest Investigators Program. This work is also partially funded by the IBEX mission as part of the NASA Explorer Program (80NSSC20K0719).

## Appendix A: Elastic Collision Differential Cross Sections

Differential cross sections for elastic collisions are calculated using the semi-classical JWKB approximation, using known interaction potentials for the considered pairs of particles. Phase shifts ($\delta_l$) are computed with the integral form given in Nitz et al. (1987, Equations A1 and A4). We calculate phase shifts for each energy until their magnitudes for ten consecutive orbital angular momentums $l$ are smaller than $10^{-5}$ radians. Scattering amplitudes are calculated for 768 values of center-of-mass scattering angles $\theta_m = \frac{\pi}{2}(1 + x_m)$, where $x_m$ is the $m$th root of the Legendre polynomial of degree 768:

$$f(\theta) = \sum_l \frac{(2l+1)}{2ik}\left(e^{2i\delta_l} - 1\right)P_l(\cos\theta), \qquad (1)$$

where $k$ is the wavenumber and $P_l$ are the Legendre polynomials. This selection of angles constitutes abscissas for the Gauss-Legendre quadrature and is the same as used in previous comprehensive studies of differential cross sections (e.g., Krstić & Schultz 1998; Schultz et al. 2016). Finally, differential cross sections are given as $d\sigma(\theta)/d\Omega = |f(\theta)|^2$.

In the case of a collision between He$^+$ ions and He atoms, we calculate separate scattering amplitudes $f^g(\theta)$ and $f^u(\theta)$ for the gerade and ungerade ground states, respectively. Moreover, the nuclei are not distinguishable and thus the charge transfer amplitude in angle $\pi - \theta$ and elastic scattering amplitude in angle $\theta$ can interfere. Consequently, the total differential cross section is

$$\frac{d\sigma}{d\Omega} = \frac{|f^g(\theta) + f^u(\theta) + f^g(\pi - \theta) - f^u(\pi - \theta)|^2}{4}, \qquad (2)$$

but elastic collisions and charge exchange (transfer) cross section can be approximately separated and expressed as:

$$\frac{d\sigma_{\text{el}}}{d\Omega} = \frac{|f^{\text{g}}(\theta) + f^{\text{u}}(\theta)|^2}{4}, \quad (3)$$

$$\frac{d\sigma_{\text{ct}}}{d\Omega} = \frac{|f^{\text{g}}(\pi - \theta) - f^{\text{u}}(\pi - \theta)|^2}{4}, \quad (4)$$

respectively. Since the elastic and charge exchange cross sections show significant peaks for angles close to 0° and 180°, respectively, the interference term accounted for in Equation (2) but neglected with the approximation in Equations (3–4) is negligible.

All considered cross sections are calculated for energies from $10^{-4}$ to $10^4$ eV. As the cross section shows significant changes for low collision energy, we calculated these cross sections with finer energy resolutions at lower energies. For ranges: $10^{-4}$–$10^{-2}$, $10^{-2}$–$10^0$, $10^0$–$10^2$, and $10^2$–$10^4$ eV, we use a logarithmically spaced grid with $\Delta(\log_{10} E) = 0.005, 0.01, 0.05$, and $0.1$, respectively. All differential cross sections calculated in this work have been posted on Zenodo under a Creative Commons Attribution license: https://doi.org/10.5281/zenodo.4555716.

## Appendix B: Monte Carlo Tracking of Elastic Collisions

Monte Carlo calculations of elastic scatterings and their consequences for the primary ISN helium population start with a test particle sample. The sample of $n = 10^6$ helium atom velocities $\{\boldsymbol{v}_{i,0}\}_{i=1,\ldots,n}$ is drawn from the Maxwell distribution with the temperature and relative speed assumed for the pristine VLISM ISN population in the global heliosphere model. Their initial position is $\{\boldsymbol{x}_{i,0}\}_{i=1,\ldots,n} = (0,0, d_0 + \delta_i)$, where $d_0 = 400$ au is set slightly farther away than the point at which the flows start to diverge from the pristine values (i.e., near the bow wave), and $\delta_i$ is randomly chosen from the range $[0, \Delta d]$, where $\Delta d$ is the calculation length step. Collisions of each particle from the sample are tracked until the particle position along the $z$-axis is closer than the heliopause at 110 au.

The probabilities of collisions are calculated with the following formula:

$$p_k = \frac{\Delta d n_k \sigma_k(v_{\text{rel},k}) v_{\text{rel},k}}{v_{\text{atom}}}, \quad (5)$$

where $n_k$ is the density of population $k$, $v_{\text{rel},k}$ is the mean relative speed between the tracked atom and this population, $\sigma_k$ is the integral cross section for the elastic collision, and $v_{\text{atom}}$ is the tracked atom speed in the Sun frame. The mean relative speed is calculated assuming that the collision partners are distributed following the Maxwell distribution (e.g., Ripken & Fahr 1983):

$$v_{\text{rel},k} = u_{\text{T},k} \left[\left(\omega + \frac{1}{2\omega}\right) \text{erf}\,\omega + \frac{\exp(-\omega^2)}{\sqrt{\pi}}\right] \quad (6)$$

with $u_{\text{T},k} = \sqrt{2k_\text{B} T_k / m_k}$, – the thermal speed and $\omega = |\boldsymbol{v}_{\text{atom}} - \boldsymbol{u}_k|/u_{\text{T},k}$ – the ratio of the relative speed of the atom $\boldsymbol{v}_{\text{atom}}$ and the bulk speed of the collision partner population $\boldsymbol{u}_k$ to the thermal speed.

The formula given in Equation (6) is precise if the integral cross section changes linearly with the relative speed. This approximation is not good for helium atom collisions with protons and He$^+$ ions, which show significant variations for low collision speeds. However, we verified that this formula could be used with a parabolic fit to the cross section in log-log scale for energies below 100 eV, which is shown with dashed lines panel (a) of Figure 1. We verified that the relative error of this approach is less than 4% for the temperature range considered in this study.

The calculation length step is set as $\Delta d = 3$ au, which assures that the probability of elastic collision with each species is much smaller than 1, so we can effectively neglect the likelihood that more than one elastic collision occurs in one length step. Additionally, this step is small enough to ignore changes in the population bulk speed and temperature. We randomly choose whether a collision with any species occurs with the probability calculated with Equation (5). The probability that no collision occurs is given as $1 - \sum_k p_k$. In this situation, the atom velocity and position are propagated:

$$\boldsymbol{v}_{i,j+1} = \boldsymbol{v}_{i,j}, \tag{7}$$

$$\boldsymbol{x}_{i,j+1} = \boldsymbol{x}_{i,j} + \Delta d \frac{\boldsymbol{v}_{i,j}}{|\boldsymbol{v}_{i,j}|}, \tag{8}$$

where $j$ enumerates positions along the atom trajectory.

If a collision with any of the considered species occurs, we draw 10 possible velocities of the collision partner from the Maxwell distribution with the speed and temperature given by the numerical model in position $\boldsymbol{x}_{i,j}$. We verified with a smaller number of modeled atom trajectories that this sample of velocities is enough to represent the difference in the collision probability with different parts of the distribution. Subsequently, we select one of these speeds using probability weights given by the integral cross section for the relative speed of the tracked atom and the collision partner. This two-step random selection provides that we account for differences in the integral cross sections with different parts of the distribution function. Further, we select the scattering angle ($\theta$) using the weighting proportional to $\sin(\theta) \, d\sigma/d\Omega$, where $d\sigma/d\Omega$ is the differential cross section for one of the two closest energy bins for which the cross section is calculated, selected with a weight inversely proportionally to the energy difference between the collision energy and the bin energy. The azimuthal angle ($\phi$) is randomly chosen from the range $[0, 2\pi]$ with uniform probability. Finally, the fractional position of the collision $\xi$ is selected from the range $[0, 1]$. The post-collision velocity and position given in this case are:

$$\boldsymbol{v}_{i,j+1} = \frac{m_{\text{atom}}}{m_{\text{atom}} + m_k} \boldsymbol{r}(\boldsymbol{v}_{i,j} - \boldsymbol{v}_{\text{CM}}, \theta, \phi) + \boldsymbol{v}_{\text{CM}}, \tag{9}$$

$$\boldsymbol{x}_{i,j+1} = \boldsymbol{x}_i + \xi \Delta d \frac{\boldsymbol{v}_{i,j}}{|\boldsymbol{v}_{i,j}|} + (1 - \xi)\Delta d \frac{\boldsymbol{v}_{i,j+1}}{|\boldsymbol{v}_{i,j+1}|}. \tag{10}$$

The rotation function $\boldsymbol{r}(\boldsymbol{v}, \theta, \phi)$ returns scattered velocity in the center-of-mass frame (moving with velocity $\boldsymbol{v}_{\text{CM}}$) and is defined in Appendix A in Swaczyna et al. (2019b). Each atom is tracked until its position along the z-axis is less than 110 au (i.e., the heliopause distance in this direction). Properties of the atom distribution are tracked in 10-au-wide bins along the z-axis, and from each trajectory, we select one speed for each bin.

## Appendix C: Asymmetric Kappa Distribution

The distribution function of the primary ISN helium atoms in the outer heliosheath is described using an asymmetric kappa distribution, in which the perpendicular components of velocities are described using a two-dimensional kappa distribution, and the parallel part is composed from two one-dimensional kappa distributions separate for the atoms with speeds below and above the distribution peak. The velocity components perpendicular to the inflow direction are denoted as $v_x$ and $v_y$, and the parallel component is along the $z$-axis: $v_z$. The perpendicular distribution function is (Livadiotis & McComas 2013):

$$f_\perp(v_x, v_y; \theta_\perp, \kappa_\perp) = \frac{1}{\pi \theta_\perp^2} \frac{\kappa_\perp - \frac{1}{2}}{\kappa_\perp - \frac{3}{2}} \left(1 + \frac{1}{\kappa_\perp - \frac{3}{2}} \frac{v_x^2 + v_y^2}{\theta_\perp^2}\right)^{-\kappa_\perp - \frac{1}{2}}, \quad (11)$$

where $\theta_\perp$ is the speed scale parameter and $\kappa_\perp$ is the kappa index in the perpendicular directions. Temperatures are related to the speed scale parameters as $\theta = \sqrt{\frac{2 k_B T}{m_{\text{He0}}}}$.

The parallel distribution function is given in the form:

$$f_\parallel(v_z; u, \theta_{\parallel S}, \theta_{\parallel A}, \kappa_{\parallel S}, \kappa_{\parallel A}) = \left(\frac{\theta_{\parallel S}}{2} \sqrt{\pi \left(\kappa_{\parallel S} - \frac{3}{2}\right)} \frac{\Gamma\left(\kappa_{\parallel S} - \frac{1}{2}\right)}{\Gamma(\kappa_{\parallel S})} + \frac{\theta_{\parallel A}}{2} \sqrt{\pi \left(\kappa_{\parallel A} - \frac{3}{2}\right)} \frac{\Gamma\left(\kappa_{\parallel A} - \frac{1}{2}\right)}{\Gamma(\kappa_{\parallel A})}\right)^{-1}$$

$$\times \begin{cases} \left(1 + \frac{1}{\kappa_{\parallel A} - \frac{3}{2}} \frac{(v_z - u)^2}{\theta_{\parallel A}^2}\right)^{-\kappa_{\parallel A}} & v_z < u \\ \left(1 + \frac{1}{\kappa_{\parallel S} - \frac{3}{2}} \frac{(v_z - u)^2}{\theta_{\parallel S}^2}\right)^{-\kappa_{\parallel S}} & v_z > u \end{cases}, \quad (12)$$

where $u$ is the peak position in the parallel direction, $\theta_{\parallel A}$ and $\theta_{\parallel S}$ are the speed scale parameters in the sunward and anti-sunward directions, and $\kappa_{\parallel A}$ and $\kappa_{\parallel S}$ are the kappa indices in these two directions. The combined distribution is given as a product of these two distribution functions:

$$f(v_x, v_y, v_z; u, \theta_\perp, \theta_{\parallel S}, \theta_{\parallel A}, \kappa_\perp, \kappa_{\parallel S}, \kappa_{\parallel A}) = f_\perp(v_x, v_y; \theta_\perp, \kappa_\perp) \times f_\parallel(v_z; u, \theta_{\parallel S}, \theta_{\parallel A}, \kappa_{\parallel S}, \kappa_{\parallel A}). \quad (13)$$

This distribution function has 7 scalar parameters. We find that the sunward part of the parallel component is consistent with the Maxwell distribution, or in other words, that $\kappa_{\parallel S}$ is very large. Therefore, we use the limit $\kappa_{\parallel S} \to \infty$ for the parallel component:

$$f_\parallel(v_z; u, \theta_{\parallel S}, \theta_{\parallel A}, \kappa_{\parallel S} \to \infty, \kappa_{\parallel A}) = \left(\frac{\theta_{\parallel S}}{2} \sqrt{\pi} + \frac{\theta_{\parallel A}}{2} \sqrt{\pi \left(\kappa_{\parallel A} - \frac{3}{2}\right)} \frac{\Gamma\left(\kappa_{\parallel A} - \frac{1}{2}\right)}{\Gamma(\kappa_{\parallel A})}\right)^{-1}$$

$$\times \begin{cases} \left(1 + \frac{1}{\kappa_{\parallel A} - \frac{3}{2}} \frac{(v_z - u)^2}{\theta_{\parallel A}^2}\right)^{-\kappa_{\parallel A}} & v_z < u \\ \exp\left(-\frac{(v_z - u)^2}{\theta_{\parallel S}^2}\right) & v_z > u \end{cases}. \quad (14)$$

We find the distribution function presented above to effectively match the numerically obtained distribution with the Monte Carlo calculations.

# References

Axford, W. I. 1972, NASSP, 308, 609

Barata, J. A. S., & Conde, C. A. N. 2010, NIMPA, 619, 21

Bzowski, M., Czechowski, A., Frisch, P. C., et al. 2019, ApJ, 882, 60

Bzowski, M., Kubiak, M. A., Czechowski, A., & Grygorczuk, J. 2017, ApJ, 845, 15

Bzowski, M., Swaczyna, P., Kubiak, M. A., et al. 2015, ApJS, 220, 28

Ceperley, D. M., & Partridge, H. 1986, JChPh, 84, 820

Chassefière, E., & Bertaux, J. L. 1987, A&A, 174, 239

Chassefière, E., Bertaux, J. L., & Sidis, V. 1986, A&A, 169, 298

Fahr, H. J. 1978, A&A, 66, 103

Fahr, H. J., Nass, H. U., & Rucinski, D. 1985, A&A, 142, 476

Frisch, P. C., Redfield, S., & Slavin, J. D. 2011, ARA&A, 49, 237

Fuselier, S. A., Bochsler, P., Chornay, D., et al. 2009, SSRv, 146, 117

Gao, R. S., Johnson, L. K., Smith, K. A., & Stebbings, R. F. 1989, PhRvA, 40, 4914

Gruntman, M. 2013, JGRA, 118, 1366

Gruntman, M. 2018, JGRA, 123, 3291

Gruntman, M. A. 1986, P&SS, 34, 387

Heerikhuisen, J., Pogorelov, N. V., Florinski, V., Zank, G. P., & Kharchenko, V. 2009, in Numerical Modeling of Space Plasma Flows: ASTRONUM-2008, ASP Conf. Ser., ed. N. V. Pogorelov, E. Audit, P. Colella, & G. P. Zank, Vol. 406 (San Francisco, CA: ASP), 189, https://ui.adsabs.harvard.edu/abs/2009ASPC..406..189H/abstract

Heerikhuisen, J., Zirnstein, E., & Pogorelov, N. 2015, JGRA, 120, 1516

Helbig, H. F., Millis, D. B., & Todd, L. W. 1970, PhRvA, 2, 771

Holzer, T. E. 1977, RvGeo, 15, 467

Izmodenov, V. V., Malama, Y. G., Kalinin, A. P., et al. 2000, Ap&SS, 274, 71

Krstić, P. S., & Schultz, D. R. 1998, Elastic and Related Transport Cross Sections for Collisions among Isotopomers of $H^+ + H$, $H^+ + H_{2M}$, $H^+ + He$, $H + H$, and $H + H_2$, Vol. 8 (Vienna, Austria: IAEA), https://www.iaea.org/publications/4710/atomic-and-plasma-material-interaction-data-for-fusion

Kubiak, M. A., Bzowski, M., Sokół, J. M., et al. 2014, ApJS, 213, 29

Kubiak, M. A., Swaczyna, P., Bzowski, M., et al. 2016, ApJS, 223, 25

Kunc, J. A., Wu, F. M., & Judge, D. L. 1983, P&SS, 31, 1157

Lee, M. A., Kucharek, H., Möbius, E., et al. 2012, ApJS, 198, 10

Lee, M. A., Möbius, E., & Leonard, T. W. 2015, ApJS, 220, 23

Livadiotis, G., & McComas, D. J. 2013, SSRv, 175, 183

Marchi, R. P., & Smith, F. T. 1965, PhRv, 139, A1025

McComas, D. J., Alexashov, D., Bzowski, M., et al. 2012, Sci, 336, 1291

McComas, D. J., Allegrini, F., Bochsler, P., et al. 2009, SSRv, 146, 11

McComas, D. J., Bzowski, M., Fuselier, S. A., et al. 2015, ApJS, 220, 22

Möbius, E., Bochsler, P., Bzowski, M., et al. 2009a, Sci, 326, 969

Möbius, E., Bzowski, M., Chalov, S., et al. 2004, A&A, 426, 897

Möbius, E., Kucharek, H., Clark, G., et al. 2009b, SSRv, 146, 149

Nitz, D. E., Gao, R. S., Johnson, L. K., Smith, K. A., & Stebbings, R. F. 1987, PhRvA, 35, 4541

Parker, E. N. 1961, ApJ, 134, 20

Pogorelov, N. V., Heerikhuisen, J., Mitchell, J. J., Cairns, I. H., & Zank, G. P. 2009, ApJ, 695, L31

Pogorelov, N. V., Heerikhuisen, J., & Zank, G. P. 2008, ApJ, 675, L41

Ripken, H. W., & Fahr, H. J. 1983, A&A, 122, 181

Schultz, D. R., Ovchinnikov, S. Y., Stancil, P. C., & Zaman, T. 2016, JPhB, 49, 084004
Schwadron, N. A., Möbius, E., Leonard, T., et al. 2015, ApJS, 220, 25
Sokół, J. M., Bzowski, M., & Tokumaru, M. 2019, ApJ, 872, 57
Sokół, J. M., Kubiak, M. A., Bzowski, M., & Swaczyna, P. 2015, ApJS, 220, 27
Swaczyna, P., Bzowski, M., Kubiak, M. A., et al. 2015, ApJS, 220, 26
Swaczyna, P., Bzowski, M., Kubiak, M. A., et al. 2018, ApJ, 854, 119
Swaczyna, P., McComas, D. J., & Schwadron, N. A. 2019a, ApJ, 871, 254
Swaczyna, P., McComas, D. J., Zirnstein, E. J., & Heerikhuisen, J. 2019b, ApJ, 887, 223
Wallis, M. K. 1975, P&SS, 23, 419
Wood, B. E., Müller, H.-R., & Möbius, E. 2019, ApJ, 881, 55
Zank, G. P. 1999, SSRv, 89, 413
Zirnstein, E. J., Heerikhuisen, J., Funsten, H. O., et al. 2016, ApJL, 818, L18